\documentclass[aps,prd,twocolumn,superscriptaddress,showpacs,bibnotes,preprintnumbers,amsmath,amssymb]{revtex4}
\usepackage{graphicx}
\usepackage{dcolumn}
\usepackage{bm}
\usepackage{epstopdf}
 \usepackage{natbib}
 \usepackage{captcont}

\IfFileExists{srcltx.sty}{\usepackage[active]{srcltx}}

\def\apj{Astrophys. J.}
\def\apjs{Astrophys. J. Suppl. Ser.}
\def\prd{Phys. Rev. D}
\def\mnras{Mon. Not. R. Astron. Soc.}
\def\jcap{J. Cos. Astropart. Phys.}
\def\aap{Astron. Astrophys.}
\def\araa{Ann. Rev. Astron. Astrophys.}

\def\apjl{Astrophys. J. Letter}

\begin{document}
\title{A Monte Carlo Bayesian Search for the Plausible Source of the Telescope Array Hotspot}
\affiliation{Department of Physics and Astronomy, University of California, Los Angeles, CA 90095-1547, USA}
\affiliation{Key Laboratory of Dark Matter and Space Astronomy, Purple Mountain Observatory, Chinese Academy of Sciences, Nanjing 210008, China }
\affiliation{Kavli IPMU (WPI), University of Tokyo, Kashiwa, Chiba 277-8568, Japan}
\affiliation{Astrophysical Big Bang Laboratory, RIKEN, Wako, Saitama 351-0198, Japan}
\affiliation{Center for Space Plasma and Aeronomic Research (CSPAR), University of Alabama in Huntsville, Huntsville, AL 35899, USA}
\affiliation{Instituto de Astrofisica de Andaluca (IAA-CSIC), P.O. Box 03004, E-18080 Granada, Spain}
\affiliation{Max-Planck-Institut f{\"u}r Kernphysik, P.O. Box 103980, 69029 Heidelberg, Germany}

\author{Hao-Ning He} 
\affiliation{Department of Physics and Astronomy, University of California, Los Angeles, CA 90095-1547, USA}
\affiliation{Key Laboratory of Dark Matter and Space Astronomy, Purple Mountain Observatory, Chinese Academy of Sciences, Nanjing 210008, China }

\author{Alexander Kusenko}
 \affiliation{Department of Physics and Astronomy, University of California, Los Angeles, CA 90095-1547, USA}
 \affiliation{Kavli IPMU (WPI), University of Tokyo, Kashiwa, Chiba 277-8568, Japan}
 
 \author{Shigehiro Nagataki}
\affiliation{Astrophysical Big Bang Laboratory, RIKEN, Wako, Saitama, Japan}
 
  \author{Bin-Bin Zhang}
\affiliation{Center for Space Plasma and Aeronomic Research (CSPAR), University of Alabama in Huntsville, Huntsville, AL 35899, USA}
\affiliation{Instituto de Astrofisica de Andaluca (IAA-CSIC), P.O. Box 03004, E-18080 Granada, Spain}
 
 \author{Rui-Zhi Yang}
\affiliation{Max-Planck-Institut f{\"u}r Kernphysik, P.O. Box 103980, 69029 Heidelberg, Germany}
\affiliation{Key Laboratory of Dark Matter and Space Astronomy, Purple Mountain Observatory, Chinese Academy of Sciences, Nanjing 210008, China }

\author{Yi-Zhong Fan}
\affiliation{Key Laboratory of Dark Matter and Space Astronomy, Purple Mountain Observatory, Chinese Academy of Sciences, Nanjing 210008, China }

\begin{abstract}
The Telescope Array (TA) collaboration has reported a hotspot of 19 ultrahigh-energy cosmic rays (UHECRs). Using a universal model with one source and energy-dependent magnetic deflections, we show that  the distribution of the TA hotspot events is consistent with a single source hypothesis, although multiple sources cannot be ruled out.  The chance probability of this distribution arising from a homogeneous distribution is $0.2\%$. We describe a Monte Carlo Bayesian (MCB) inference approach, 
which can be used to derive parameters of the magnetic fields as well as the source coordinates, and we apply this method to the TA hotspot data, inferring the location of the likely source.  
We discuss possible applications of the same approach to future data. 
\end{abstract}

\maketitle

The Telescope Array (TA) collaboration has reported 72 cosmic-ray events with energies above $57~\rm EeV$\cite{TA2014}
using the surface detector (SD) data recorded between May 11, 2008 and May 4, 2013.
The data show a {\em hotspot}, 19 events
clustered in a circle of $20^\circ$ radius centered at
${\rm R.A.}=146.^\circ7$, ${\rm Dec.}=43.^\circ2$ in the equatorial coordinates.
Understanding the origin of this hotspot can shed new light on sources of ultra-high-energy cosmic rays (UHECRs).

The origin of UHECRs remains unknown~\cite{Bhattacharjee:1998qc}.
Active galactic nuclei (AGN)~\cite{Murase2012} and gamma-ray bursts (GRBs)\cite{Waxman1995} 
are among the likely sources of extragalactic UHECRs,
while the galactic hypernovae and GRBs that occurred in the past may be responsible 
for some fraction of UHECRs~\cite{Calvez:2010uh}.
Alternatively, fast rotating magnetars~\cite{Arons2003}, newly-born pulsars \cite{Fang2012} 
or young pulsar winds \cite{Lemoine2014} are also source candidates of UHECRs.
Additionally, galaxy clusters~\cite{Norman1995} and starburst galaxies with strong galactic wind~\cite{Anchordoqui2001} 
are suggested to be able to accelerate particles to ultra-high energy. 
Anisotropies in arrival directions of UHECRs and their temporal distributions can help identify their sources. However, since charged particles are deflected by magnetic fields, it is important to take into account the effects of such deflections on the arrival directions and temporal distributions of UHECRs.

The TA hotspot is not the only peculiarity in the UHECR data. An excess of UHECR events around the nearby radio galaxy Centaurus A was observed by the Pierre Auger Observatory (PAO)~\cite{PAO2008}.
\citet{Yuksel2012} assumed that Centaurus A was the source of the excess, they 
and used the angular distribution of the excess events to constrain the extragalactic magnetic fields. \citet{Takami2012} and \citet{Farrar2013} examined the possible  contribution of Centaurus A to the observed hotspot by simulating the propagation of UHECR protons and nuclei in the magnetic field. 

In what follows we will focus on the TA hotspot, and we will assume that this excess is due to a single source.  

Let us consider the temporal information in the TA hotspot data and identify the types of  sources that can be consistent with it.  Particles deflected in the magnetic field with a deflection angle $\theta$ arrived at Earth later than photons propagating rectilinearly,
with an average time delay \cite{Finley2006} 
\begin{equation}
\Delta T=3.3\times10^6{\rm yr}\ \frac{D}{1\ \rm Mpc}\left(\frac{\theta}{\sin\theta}-1\right).
\end{equation}
The distribution of the arrival times has a standard deviation $\sigma_d\sim \Delta T$~\cite{Waxman1996}, where $\left(\frac{\theta}{\sin\theta}-1\right)\approx 0.02$ for $\theta\sim 20^\circ$.
This should be juxtaposed with the fact that the 19 cosmic rays are observed within a time window of 5 years.

Besides a source active on some long time scales, it is reasonable to consider transient sources which satisfy the constraints on the energy budget and the rate. 
If the 19 events observed over a time $T_{\rm obs}=5$~years were emitted from a single  short burst, then they represent a fraction $\eta_{\rm max}\sim (\Delta T/T_{\rm obs}) $ of the total burst energy.  While the emission of cosmic rays is likely to be beamed, it is useful to determine the isotropic equivalent energy 
emitted in cosmic rays with energies above $57$~EeV:
\begin{align}
E_{>57\rm EeV}& =  
\left ( F_{\rm hs} \Omega_{\rm hs}T_{\rm obs} D^2 \right ) 
\left (\frac{4\pi}{\Omega_{\rm jet}} \right ) \nonumber \\ & = 8\times 10^{52}\ 
{\rm erg}\left(\frac{D}{1\ \rm Mpc}\right)^2
\end{align}
Here $\Omega_{\rm hs}$ and $\Omega_{\rm jet}$ are the solid angles of the hotspot and the jet, which are 0.38 and 0.03, respectively.  
The observed flux in the hotspot~\cite{Fang2014} is 
$F_{\rm hs}\simeq(4.4\pm1.0)\times10^{-11}\rm erg~s^{-1} ~cm^{-2}~sr^{-1}$. 
Assuming an injected cosmic ray spectrum of $\frac{dN}{dE}\sim E^{-2}$,
the total injected energy of cosmic rays which cover 12 decades of energy is about
$50$ times the injected energy in the range of $60-100$ EeV.
Then the requisite total isotropic injected energy of the single transient source is $4\times 10^{54}{\rm erg}\left(\frac{D}{1\ \rm Mpc}\right)^2$. 
Therefore, a GRB with an extremely high kinetic energy at a distance $\sim$~Mpc could produce the observed hotspot.  Although no such nearby event has been 
observed so far~\cite{Fan2013}, it is possible that such a GRB took place during the $10^5$~years time window.

On the other hand, if the 19 cosmic rays are contributed by multiple transient sources in a single galaxy, 
there's no constraint on the energy budget, but a constraint on the rate of the transients in the galaxy. A single star-forming galaxy hosting several GRBs during a time period of $\sim6.6\times10^4{\rm yr}\frac{D}{1\rm Mpc}$, could be the source of the hotspot.  Let us assume that each transient source contributed one event.  Taking into account a beaming correction for the GRB rate, which is a factor of $75\pm 20$~\cite{Piran2005}, 
the GRB rate in the star-forming galaxy should exceed  $(0.04\pm0.01)\left(\frac{D}{1\rm Mpc}\right)^{-1}\rm yr^{-1}$ per galaxy.
The supernova (SN) rate $R_{\rm SN}$ correlates with the star formation rate (SFR), $R_{\rm SN}=1.2\times10^{-2}{\rm yr^{-1}}\frac{{\rm SFR}}{M_\odot \rm yr^{-1}}$~\cite{Fukugita2003}.
Adopting the observed ratio of GRB rate to SN rate $R_{\rm GRB}/R_{\rm SN}=0.5-4\%$ \cite{DellaValle2006},
one can derive GRB rate as a function of SFR.
Then combining the correlation between SFR and the far-infrared luminosity $L_{\rm FIR}$,
${\rm SFR}=1.71 M_\odot {\rm yr^{-1}}\frac{L_{\rm FIR}}{10^{10}L_{\odot}}$\cite{Kennicutt1998},
one can estimate the correlation between GRB rate and $L_{\rm FIR}$ as 
$R_{\rm GRB}=(1-8)\times10^{-4}{ \rm yr^{-1}}\frac{L_{\rm FIR}}{10^{10}L_{\odot}}$.
Therefore, a star-forming galaxy with 
\begin{equation}\label{LFIR}
\frac{L_{\rm FIR}}{10^{10}L_{\odot}}> 400\left(\frac{D}{1\rm Mpc}\right)^{-1}
\end{equation}
can be a possible source.  This implies that only a fraction of star-forming galaxies produce hotspot events~\cite{Magnelli2011}.

Let us now consider propagation of cosmic rays from a single source within 200~Mpc, from which at least $10\%$ of UHECRs with relevant energies can
reach Earth without a significant energy attenuation~\cite{Ave2002}.
Since the high-energy gamma-ray emission indicate an extreme particle acceleration and large energy conversions,
we search for sources among the long-term active sources from Fermi LAT catalogue and TeV catalogue (http://tevcat.uchicago.edu), 
including massive galaxy clusters~\cite{Ackermann2014}, BL Lac objects \cite{Ackermann2015},
radio galaxies and starburst galaxies~\cite{Nolan2012} with high-energy gamma-ray observations,  
plus star-forming galaxies satisfying the criteria in Eq.~(\ref{LFIR})~\cite{Ackermann2012}. 

To describe the effect of the magnetic field on the cosmic rays from a single source, we separate the magnetic fields into a regular component and a random component. The regular magnetic field deflects cosmic rays in the same direction, and the deflection angle is inversely proportional to the magnetic rigidity~\cite{Finley2006}:
\begin{equation}\label{eq_reg}
\delta_{\rm reg}\simeq 0.5^\circ Z\frac{100\, {\rm EeV}}{E}\frac{D_{\rm reg}}{1\rm Mpc}\frac{B_{{\rm reg},\perp}}{1\rm nG}=A_1\times \frac{100\, {\rm EeV}}{E},
\end{equation}
where $B_{{\rm reg},\perp}$ is the strength of magnetic field perpendicular to the propagation path,
$D_{\rm reg}$ is the propagation length in the regular magnetic field, and $Z$ is the charge of the nucleus.  Here we have defined a parameter 
$A_1=0.5^\circ Z\frac{D_{\rm reg}}{1\rm Mpc}\frac{B_{{\rm reg,}\perp}}{1\rm nG}$ that depends on the magnetic field and composition of cosmic rays. 

The random magnetic field can be treated as a collection of domains with randomly oriented magnetic fields, which cause the cosmic rays to perform a random walk. The distribution of the deflection angles $\delta_{\rm dif}$ follows a Gaussian distribution:
\begin{equation}\label{eq_gauss}
f(\delta_{\rm dif},\delta_{\rm rms})=\frac{1}{\delta_{\rm rms}\sqrt{2\pi}}\exp\left(-\frac{\delta_{\rm dif}^2}{2\delta_{\rm rms}^2}\right)
\end{equation}
The root mean squared (rms) deflection angles of particles are inversely proportional to the magnetic rigidity~\cite{Harari2002b}:
\begin{eqnarray}\label{eq_dif}
\delta_{\rm rms}&\simeq & 0.36^{\circ}Z  \frac{100\, \rm EeV}{E} \left(\frac{D_{\rm dif}}{1{\rm Mpc}}\right)^{\frac{1}{2}}
\left(\frac{D_{\rm c}}{1{\rm Mpc}}\right)^{\frac{1}{2}}\frac{B_{\rm rms}}{1~{\rm nG}}\nonumber\\
&=& A_2\times  \frac{100\, \rm EeV}{E},
\end{eqnarray}
where $B_{\rm rms}$ and $D_{\rm c}$ are the rms strength and the coherence length of the random magnetic field, 
$D_{\rm dif}$ is the propagation length in the random magnetic field, 
and $A_2=0.36^{\circ}Z \left(\frac{D_{\rm dif}}{1{\rm Mpc}}\right)^{\frac{1}{2}} \left(\frac{D_{\rm c}}{1{\rm Mpc}}\right)^{\frac{1}{2}}\frac{B_{\rm rms}}{1~{\rm nG}}$ 
is a parameter depending on the features of the random magnetic field, the propagation path, and the composition of cosmic rays.
We note here Eq. (\ref{eq_dif}) corresponds to the small deflections of cosmic rays when the coherence length is much larger than the projected transverse deflection, and the propagation length of cosmic rays is larger than the maximum turbulence scale.
\footnote{We note that if the observed cosmic rays form multiple images of the same source and $\delta_{\rm dif}\sim D_{\rm c}/D_{\rm dif}$ or $\delta_{\rm dif}> few~D_{\rm c}/D_{\rm dif}$, then  Eq.~(\ref{eq_dif}) is not valid~\cite{Harari2002b,Farrar2014}.}

\begin{figure}[ht]
\includegraphics[width=85mm,angle=0]{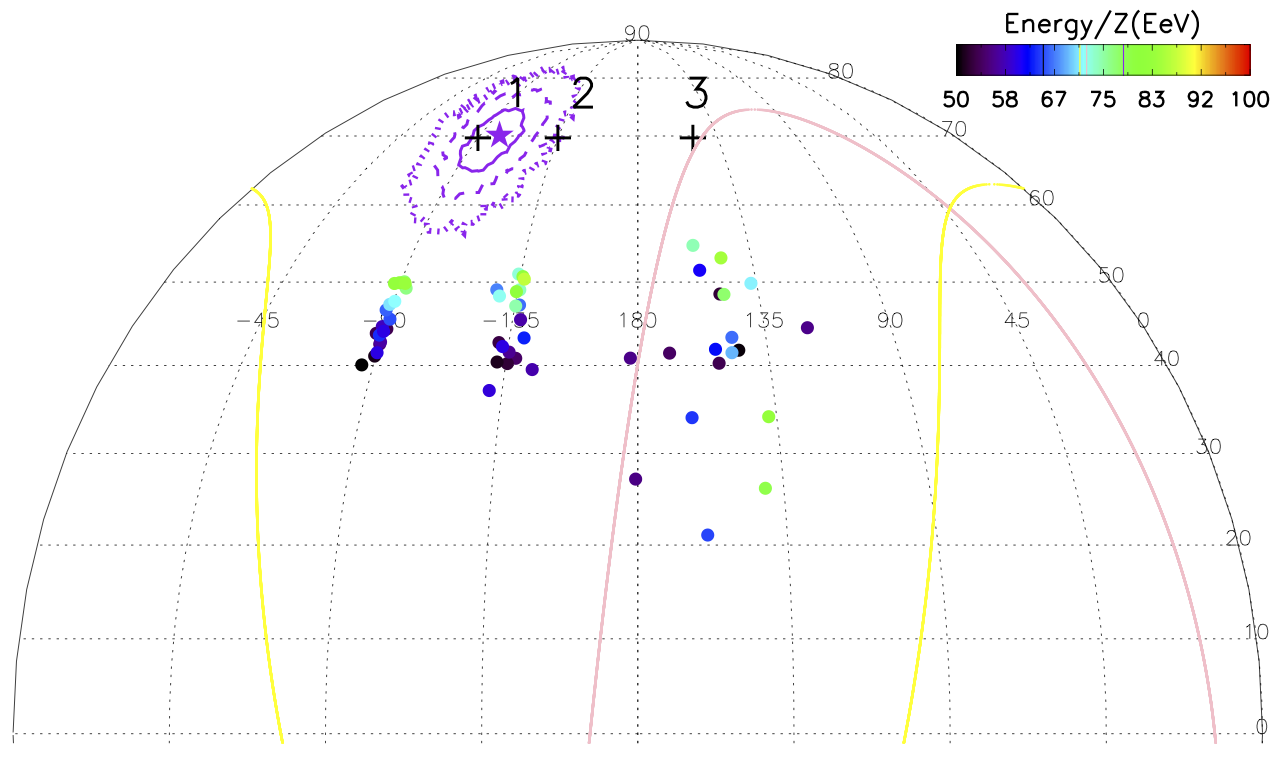}
\caption{ Three simulated clusters of 19 cosmic rays each (filled circles) and the corresponding sources 
(labelled 1, 2 and 3) in the equatorial coordinates.  These three simulations correspond to $ A_2=$1, 3, and 10, respectively.  
The best-fit reconstructed source of the first group of events is marked by the purple star symbol, and its 1$\sigma$, 2$\sigma$, and 3$\sigma$ error contours are denoted by the purple solid, dashed and dotted curves, respectively.
The yellow and pink lines represent the galactic plane and the SGP, respectively. 
} \label{3cases}
\end{figure}

If the particles of the same energy and rigidity are emitted from a point source, and if they propagate only in the regular magnetic field,
the magnetic deflections produce an apparent shift in the position of the source. The shifted sources of cosmic rays with different rigidities are aligned with the original source, and the ones with the lower rigidity lie farther from the original source. In that case, one can locate the original source as a point along the line of shifted sources
by taking into account the dependencies of shift angles on the magnetic rigidity.
However, the diffusion effects in the random magnetic field turn shifted point sources into diffuse patches.  The effects of diffusion are characterized by the parameter $A_2$.  Three simulated hotspots for $A_2=[1, 3, 10]$ from left to right, are plotted in Fig.\ref{3cases},
and marked as weakly, mildly and strongly diffused hotspots, respectively. 
The events in the weakly diffused hotspot form a line pointing to the source.

The size of the observed TA hotspot is similar to the size of the strongly diffused hotspot. 
We assume the composition of the TA hotspot events is pure, and it can be described by a single charge $Z$.
To calculate the chance probability of the TA hotspot with acceptable statistics, 
we divide the events into two energy bins: $E>75~\rm EeV$ 
and $E<75~\rm EeV$, then obtain two diffuse images, of which the one with a lower energy has a larger footprint.
As in Eq. (\ref{eq_dif}), the parameter $A_2$
can be measured as the product of $\delta_{\rm rms}$ and the average energy (in unit of $100~\rm EeV$).
For the two groups with $\delta_{\rm rms}=(10.7^\circ, 14.1^\circ)$ corresponding to the average energies $E=(82.0~{\rm EeV}, 62.7~{\rm EeV})$,
the resulting $A_2$ are compatible,
which is consistent with the hypothesis that the TA hotspot events are from a single source.
We call the hotspot with the above structure  as a magnetic-selected hotspot.

\begin{figure}[ht]
\includegraphics[width=85mm,angle=0]{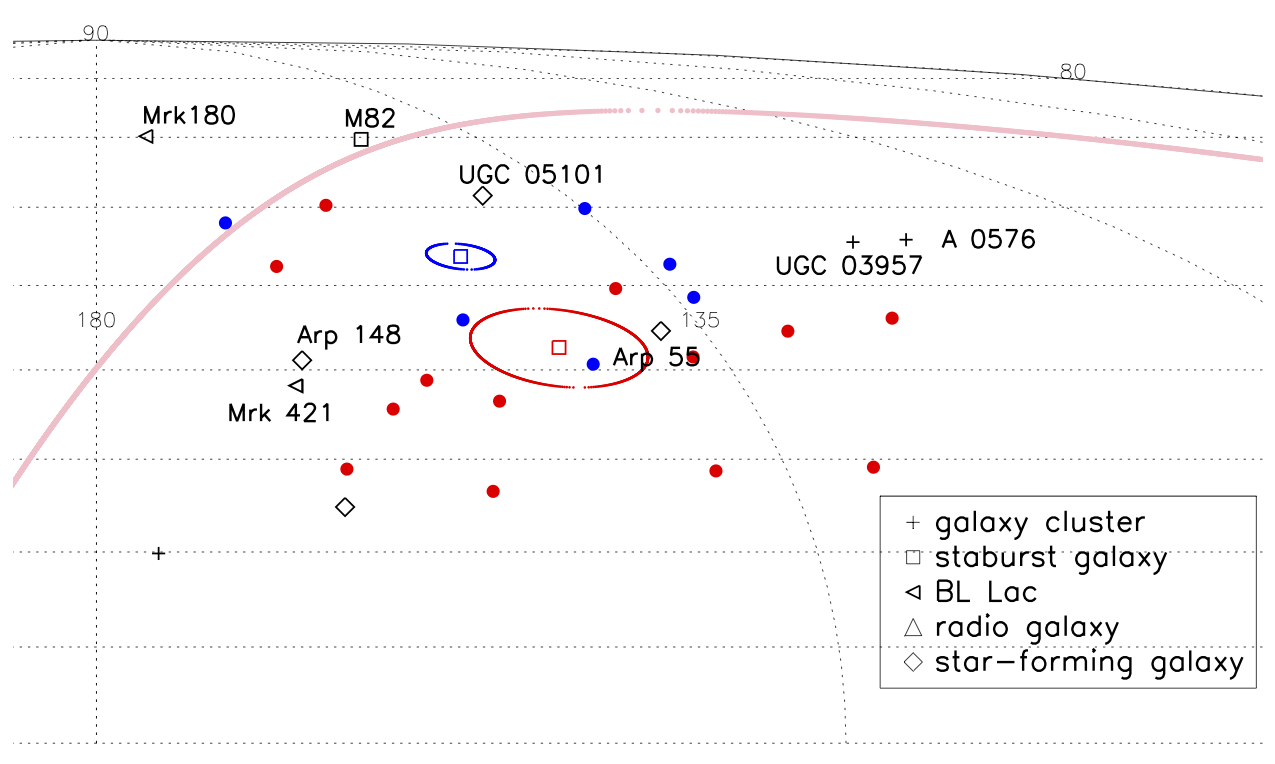}
\caption{ The 19 events at the hotspot in the equatorial coordinates are denoted by filled circles ({\it red:} $E< 75{\rm EeV}$; {\it blue:} $E > 75\rm{EeV}$).  Reconstructed positions of shifted sources for two groups of the hotspot events are denoted by the open squares; the errors are shown by ellipses of the corresponding color. 
} \label{bin}
\end{figure}

We generated $18,000$ realizations of 72 events with the same energy as the observed ones,
randomly distributed following the TA exposure, to simulate the TA observations. 
Since the detection efficiency above $57$ EeV is $\sim 100\%$,
we adopt a geometrical exposure $g(\theta)=\sin\theta\cos\theta$ depending on zenith angle $\theta$.
The observatory located at ($39^\circ.30$N, $112^\circ.91W$), 
one can derive the exposure of TA observatory as a function of declination 
via integrating the geometrical exposure over the time from May 11, 2008 to May 4, 2013.  
Then we randomly distribute 72 events on the sky from $0^\circ$ to $360^\circ$ in R.A. and $-15^\circ$ to $90^\circ$ in DeC,
according to the TA exposure.

Out of $18,000$ realizations, there are $37$ realizations with a hotspot with more than 19 events in $20^\circ$-radius circle exists,
composed by two clusters in two energy bins ($E<75~\rm EeV$ and $E>75~\rm EeV$) with the amount of events larger than 13 and 6, 
each of which has the rms of diffuse angle smaller than $10.7^\circ$ and $14.1^\circ$, 
respectively.
This suggests that 0.2$\%$ random realizations would produce a similar hotspot detection,
implying the probability of $99.8\%$ that the magnetic-selected structure of the TA hotspot is not from a fluctuation.

The positions of the shifted sources, identified as the spacial center of the group of events, are shown in Fig.~\ref{bin}.
In principle, one can locate the original source because the line connecting the two shifted sources should lead to the true source located at  
the distance 
$\delta_{\rm lo}=\frac{ E_{\rm hi}}{E_{\rm hi}-E_{\rm lo}}\Delta\delta=4.3\ \Delta\delta$, from the low-energy shifted source. Here 
 $\Delta\delta$ is the separation angle between the two shifted sources, and 
 $E_{\rm hi}$($E_{\rm lo}$) and $\delta_{\rm hi}$($\delta_{\rm lo}$) are the average energy and the shift angle of the high-energy (low-energy) group of events. However, in practice this approach is stymied by low statistics. To quantitatively calculate the probabilities of the source candidates,
we employ the Monte Carlo Bayesian (MCB) inference approach~\cite{zhangbb15}.

Let us assume that the source is at ($\rm R.A., Dec.$) and the magnetic field is described by three parameters ($\alpha, A_1,A_2$), 
where $\alpha$ is the clockwise angle between the direction to the north pole.
From the coordinate of the original source and parameters $\alpha$ and $A_1$,  
for each event with different energy,
one can derive the coordinate of the corresponding shifted source.
Furthermore, one can measure the diffusion angle $\delta_{{\rm dif},i}({\rm R.A.,Dec.},\alpha,A_1,E_{i})$ between the $i-th$ event and its shifted source.
The probability of a parameter set for the $i-th$ event can be calculated via $f_i(\delta_{{\rm dif},i}({\rm R.A.,Dec.},\alpha,A_1,E_i), \delta_{{\rm rms},i}(A_2,E_i))$
according to  Eq.~(\ref{eq_gauss}) and Eq.~(\ref{eq_dif}).
Therefore, the probability of a parameter set for the cluster of events with the amount of $N$ is
\begin{equation}
P\propto\prod\limits^{N}_{i=0} f_i(\delta_{{\rm dif},i}({\rm R.A.,Dec.},\alpha,A_1,E_i), \delta_{{\rm rms},i}(A_2,E_i)).
\end{equation}
We then define the log-likelihood function as 
\begin{eqnarray}
\label{like}
L
&\equiv&ln(P)\nonumber\\
&=&\sum\limits^{N}_{i=0} ln(f_i(\delta_{{\rm dif},i}({\rm R.A.,Dec.},\alpha,A_1,E_i), \delta_{{\rm rms},i}(A_2,E_i))\nonumber\\
&&+{\rm const}.
\end{eqnarray}

\begin{figure}[ht]
\includegraphics[width=85mm,angle=0]{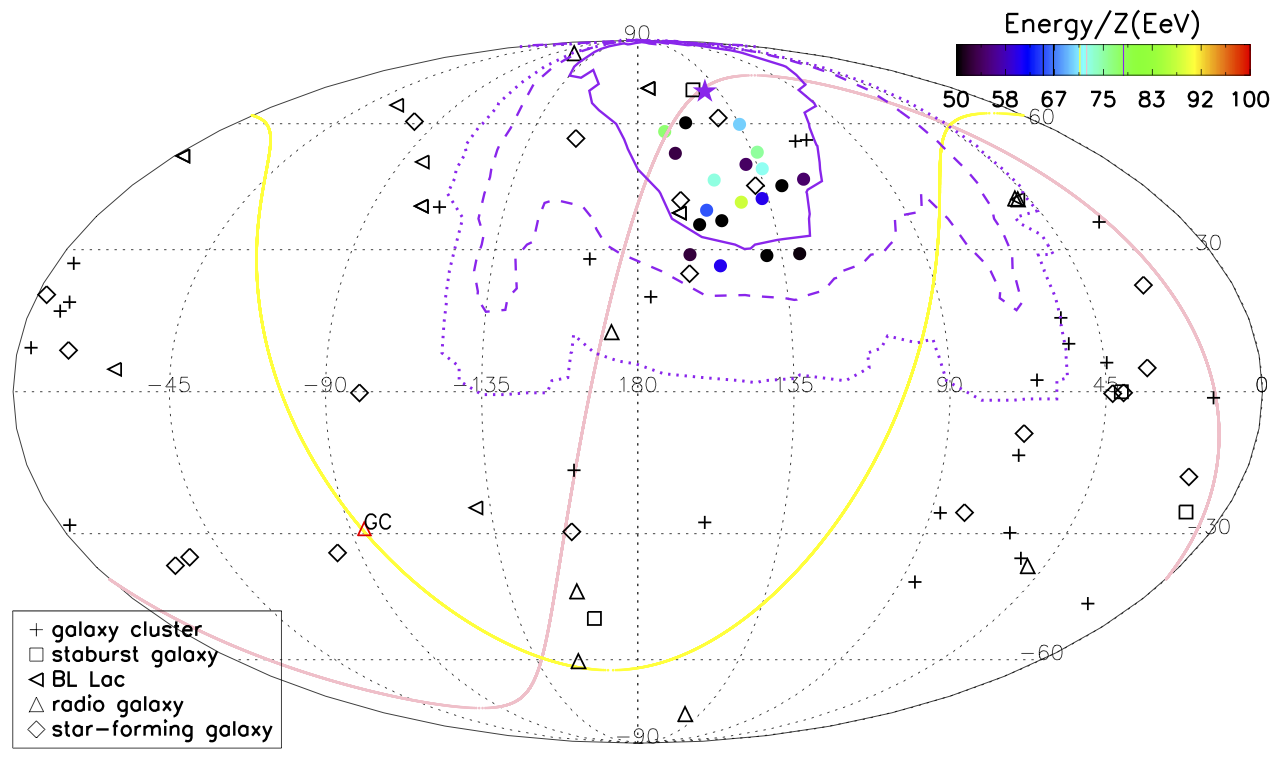}
\caption{ The 19 events (filled colorize circles) at the hotspot in the equatorial coordinates.
The purple star represent the best-fit coordinate, and
the solid, dashed and dotted purple lines represent the 1$\sigma$, 2$\sigma$, 3$\sigma$ contours for the source coordinates, 
for our 5-parameter MCB calculation.
} \label{fig1}
\end{figure}

\begin{figure}[ht]
\includegraphics[width=85mm,angle=0]{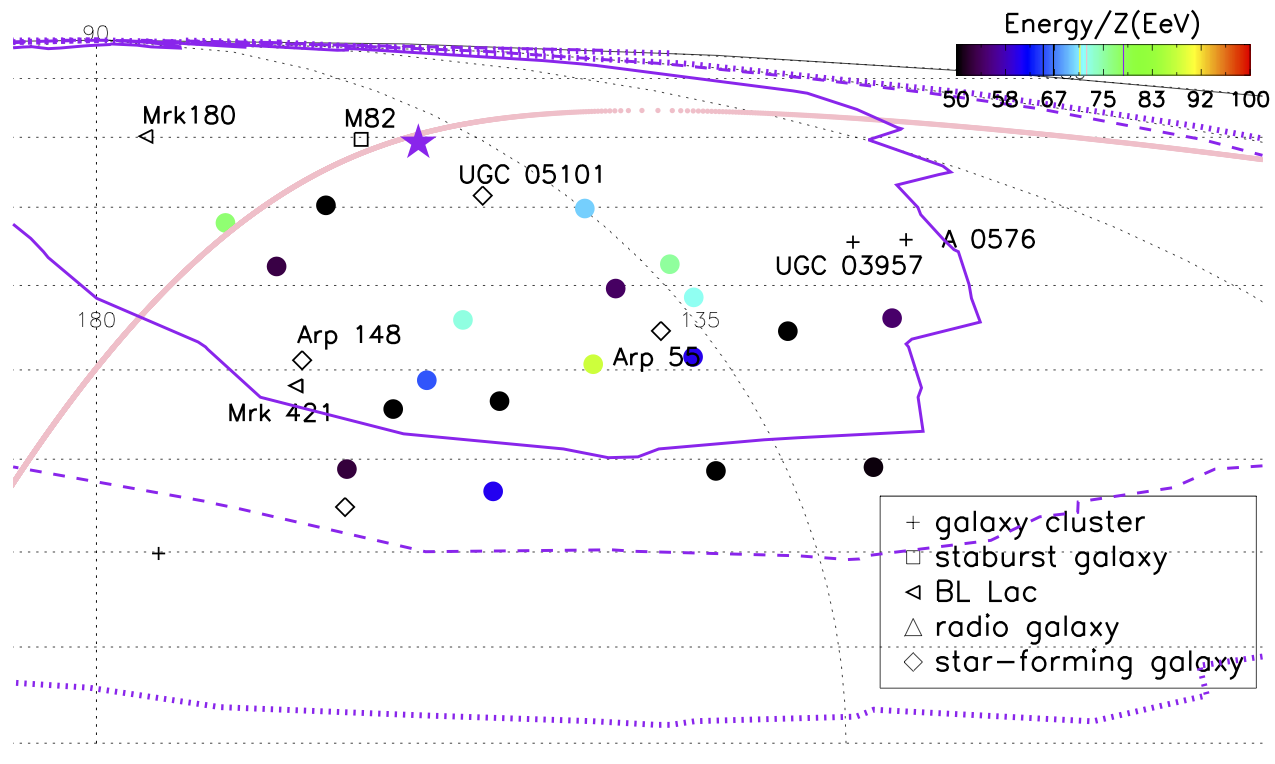}
\caption{ A enlarged plot of the hotspot region. 
} \label{fig2}
\end{figure}

Using the log-likelihood function Eq. (\ref{like}) in our Monte Carlo (MC) fitting engine~\cite{zhangbb15}, 
the best-fit parameters and their uncertainties can be realistically determined by the converged MC chains. 
For a weakly diffused cluster, such as the one from Source 1 in Fig. \ref{3cases},  
our MCB method will derive small error contours of the source position, as shown by the purple curves in Fig. \ref{3cases},
which can well locate the source.
However, for the strongly diffused hotspot, 
such as the TA observed hotspot in Fig. \ref{fig1},
our method will get a larger error contours of the sources position with the current statistics,
which might include other sources besides the true source,
as in Figs.~\ref{fig1} and \ref{fig2}.

The best-fit 5 parameters with 1-sigma uncertainties derived by the MCB approach of the TA observed hotspot are listed in Table~\ref{tab3}.
We note that the starburst galaxy M82 is close to the best-fit source position, denoted by the purple star symbol in Figs.~\ref{fig1} and \ref{fig2}. 
There are 7 other objects within the 1$-\sigma$ error contour, with their probabilities relative to the largest probability of the best-fit case listed in Table~\ref{tab3}. 
The starburst galaxy M82 has the largest relative probability $99.8\%$, 
and the star-forming galaxy UGC 05101 and the blazar Mrk 180 also have comparably large relative probabilities of being the source.
The probabilities are posterior ones.

\begin{table*}[ht!]
\begin{center}
\begin{tabular}{ccccccccccc}
\hline
Source Name&Source Type&Distance&RA& Dec&$\alpha$&$A_1$&$A_2$&${P}/{P_{\rm bes-fit}}$\\
&&(Mpc)&($^\circ$)&($^\circ$)&($^\circ$)&($^\circ$)&($^\circ$)&(\%)\\
\hline
\hline
best-fit&-&-&$142.8_{-40.0}^{+47.6}$&$69.2_{-27.6}^{+11.7}$&$185.7_{-121.2}^{+109.6}$&$17.4_{11.0}^{+17.0}$&$9.4_{-0.3}^{+3.7}$&$100$\\
M82&starburst galaxy&3.4&149.0&69.7&174.2&17.6&9.6&99.8\\
UGC 05101&star-forming galaxy&160.2&143.0&61.5&182.9&11.6&9.2&96.9\\
Mrk 180&blazar&185&174.1&70.2&136.1&19.9&9.3&91.3\\
UGC 03957&galaxy cluster&150.3&115.2&55.4&253.4&14.9&9.5&67.4\\ 
A 0576&galaxy cluster&169.0&110.4&55.7&259.0&17.0&9.4&$63.4$\\
Arp 55&star-forming Galaxy&162.7&138.2&44.5&279.6&1.9&9.7&$55.3$\\
Arp 148&star-forming Galaxy&143.3&165.3&41.1&69.3&10.5&10.0&$41.8$\\
Mrk 421&blazar&134&166.1&38.2&61.5&11.2&9.9&$35.6$\\
\hline
\end{tabular}
\end{center}
\caption{The best-fit parameters (with 1-$\sigma$ errors) are juxtaposed with 
 8 source candidates.  The probability of being the source of the TA hotspot is listed for each candidate in the last column.
 }
\label{tab3}
\end{table*}

Due to the poor statistics of the current data, one cannot exclude any of the source candidates in Table~\ref{tab3}.

 Sub-PeV to PeV neutrinos can also be used to identify the sources.
Three out of 54 IceCube neutrino events, namely events 9, 27,and 50, overlap with the location of the hotspot \cite{Fang2014,Kopper2015}.
Although the present dataset of IceCube neutrinos does not provide a sufficient discriminating power,
M82 can be a good candidate source of neutrinos since it is very close to Earth, 
and it can probably accelerate cosmic rays to ultrahigh energies.
For example, particles can be accelerated by supernovae, 
which are plentiful in this starburst galaxy with star formation rate 10 times higher than that of the Milky Way.  The supernova shocks can accelerate cosmic rays to the energy as high as ~PeV, and they can be further accelerated by the supergalactic wind to ultra-high energy~\cite{Anchordoqui2001}.
The superwind kinetic energy $\dot{E}_{\rm sw}\sim 2.7\times10^{42}\rm erg~s^{-1}$, 
implies that M82 is energetic enough to produce the flux of UHECRs in the hotspot~\cite{Anchordoqui2001}.
Compared to M82, the blazar Mrk~180 is much further from Earth,
the observed UHECRs are suppression by a factor of $95\%$ \cite{Ave2002}.
The flux of TeV gamma-ray emission from Mrk~180 is as high as 0.11 crab~\cite{Mrk1802006} at $200$~GeV.
Assuming a hadronic model~\cite{Murase2012} and the accelerated cosmic ray spectrum index $\approx 2$,
Mrk~180 can also be energetic enough to produce UHECRs at the hotspot.
The star-forming galaxy UGC 05101 has a high far-infrared luminosity $L_{\rm FIR }=89\times10^{10}L_\odot$.
According to \citet{He2013}, assuming a half-light radius of $\sim1~\rm kpc$, 
the energy-loss time is $\tau_{\rm loss}=1.8\times 10^{4}{\rm yr}\frac{l}{100\rm pc}$ 
with $l$ as the scale of the dense region in the galaxy,
and the confinement time is $\tau_{\rm conf}=2.7\times10^3{\rm yr}\left(\frac{E_{\rm p}}{50\rm EeV}\right)^{-0.5}$,
which is shorter than the energy-loss time for ultra-high energy cosmic rays,
implying that ultra-high energy cosmic rays can escape from UGC 05101.

Future more GeV-TeV $\gamma-$ray detections by the Fermi Large Area Telescope(Fermi-LAT)~\cite{Fermi2015}, 
High Energy Stereoscopic System (H.E.S.S.)\cite{HESS1997}, the High Altitude Water
Cherenkov detector array (HAWC)~\cite{HAWC2013}, the Very Energetic Radiation Imaging Telescope Array System (VERITAS)\cite{VERITAS2015},
the Large High Altitude Air Shower Observatory (LHAASO)~\cite{Cao2010}  and
the Cherenkov Telescope Array (CTA)~\cite{CTA2011} 
on sources can provide more hints on UHECR acceleration of those source candidates~\cite{Murase2012}.

One possible way to distinguish among those sources is to check 
whether the spectrum of the hotspot events exhibits a GZK cutoff. 
The cutoff is expected in the spectrum except for the nearby source M82.
However, it is not possible so far to determine definitively whether the GZK suppression is present in the hotspot spectrum for the current statistics.
Another way to distinguish the sources is to observe GZK neutrinos~\cite{Stecker:1978ah}.
No GZK neutrinos can be produced by UHECRs from M82 traveling a short distance.
For the other sources with distance about $200$ Mpc, assuming a pure proton composition,
$90\%$ UHECRs are strongly attenuated by photo-meson interactions with CMB photons,
and produce neutrinos.  Based on the observed flux of UHECRs at the hotspot\cite{Fang2014}, 
we estimate the flux of EeV neutrinos to be $\sim(2.2\pm 0.5)\times10^{-8}\rm~GeV~cm^{-2}~s^{-1}~sr^{-1}$,
which might be detected by the IceCube\cite{Kopper2015} or the PAO\cite{PAO2015} in the near future.
However, for heavy composition, the cosmic rays would lose energy through photo-disintegration, 
leading to a suppression on the flux of EeV GZK neutrinos and an enhancement on the flux of PeV neutrinos~\cite{Hooper2005}.
In addition, the high-energy cutoff may also come from the intrinsic cutoff at the source 
caused by its limited acceleration capability. In that case, the flux of GZK neutrinos would be much lower.

Therefore, all possibilities remain open at present. Future data can help resolve the ambiguity. 
If the source is confirmed in the future, one can fit the 3 relevant parameters of magnetic fields via the MCB method, as listed in Tab.~\ref{tab3},
and further constrain the magnetic field.
From Tab.~\ref{tab3}, $A_2\sim9-10$, indicates a random magnetic field with features of 
$Z\left(\frac{D_{\rm dif}}{1{\rm Mpc}}\right)^{\frac{1}{2}} \left(\frac{D_{\rm c}}{1{\rm Mpc}}\right)^{\frac{1}{2}}\frac{B_{\rm rms}}{1~{\rm nG}}=25-28$.
The feature of the regular magnetic fields can be described by the fitted parameters $\alpha$ and $A_1$, which are varied for different sources.
The characteristic value of $A_1=Z\frac{D_{\rm reg}}{1\rm Mpc}\frac{B_{{\rm reg},\perp}}{1\rm nG}$
ranges from $\sim 4$ to $\sim40$. 
The strength of the magnetic field, the length of the propagation path and the composition of the UHECRs are degenerate.
The large value of $A_2$ and $A_1$ may be considered as the hint of a heavy composition, a strong magnetic field or a distant source.
However, we cannot remove the degeneracy between the three factors due to current poor statistics.

\begin{figure}[ht]
\includegraphics[width=85mm,angle=0]{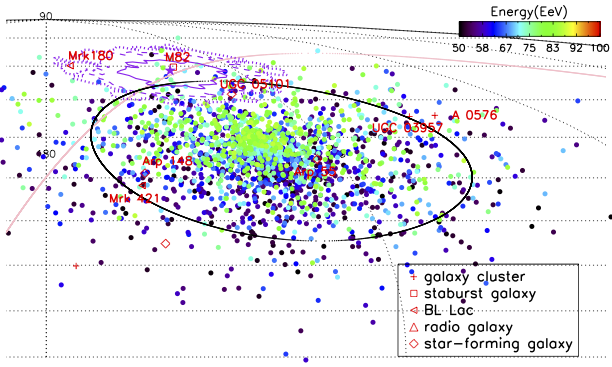}
\caption{ A enlarged plot of the simulated data from M82 (colorized filled circles).
The black circle is the 20$^\circ$-radius circle with the maximum amount of events. 
The solid, dashed and dotted purple lines are 1,2,3~$\sigma$ contours for source coordinates
derived from the simulated data within the black circle. The purple star denotes the best-fit source coordinates. 
} \label{fig_simulation}
\end{figure}

As the statistics increase in the future observation by TA$\times4$~\cite{Fukushima2015}
or JEM-EUSO~\cite{Olinto2015}, the error of the source position will be reduced. 
We have simulated a hotspot observation with about 2000 events originating from M82, to test whether our MCB method can trace back to the source.
The simulated data consist of cosmic rays from M82, 
with a spectrum of $\frac{dN}{dE}\propto E^{-2}$, and deflected by the regular and random magnetic fields, 
with the parameter set ($\alpha, A_1, A_2$)=(174.2,17.6, 9.6), derived from the current observations, as shown in Tab. 1,
and a uniformly random background with the signal to background ratio as $10\%$.
The probability of a background event can be described by the 6th parameter $R_{\rm b}$,
then the probability for one event can be calculated via
$f_{{\rm s},i}=f_i(\delta_{{\rm reg},i}({\rm R.A.,Dec.},\alpha,A_1), \delta_{{\rm rms},i}(A_2))+R_{\rm b}$.
We find a $20^\circ$-radius circle, within which the number of data $N_{\rm on}$ reaches the maximum value.
Then the log-likelihood function for the dataset within the circle is written as
$L_{\rm s}=\sum\limits^{N_{\rm on}}_{i=0} ln(f_{{\rm s},i})+{\rm const}$.
As in Fig. \ref{fig_simulation}, the 1~$\sigma$ contour of the possible source is narrowed down to only cover a small region around the source, 
The result is sensitive to the background level, a lower background level will lead to a smaller error.

In summary, we have explored the hypothesis of a single source for the TA hotspot using a universal model of cosmic rays from a single source  deflected by magnetic fields. We generated predictions for different types of hotspots with the different magnitudes of diffusion by the random magnetic field.
Our analysis shows that the distribution of the TA hotspot events is 
consistent with the single source hypothesis,
and the chance probability of this distribution is $0.2\%$.
The MCB method can be used to find out the best-fit source coordinates and magnetic field parameters.
This method suggests that M82, UGC~05101 and Mrk~180 are likely candidates, 
although the other sources listed in Table~\ref{tab3} cannot be excluded. 
The MCB method can be adopted to other magnetically selected hotspots that may be observed in the future.

\medskip
H.N.H. is supported by National Natural Science of China under Grants No. 11303098 and No. 11433009,
China Postdoctoral science foundation under Grants No. 2012M521137 and No. 2013T60569,
and the International Postdoctoral Exchange Fellowship Program, as well as by the U.S. Department of Energy Grant No. DE-SC0009937. 
The work of A.K.  was supported by the U.S. Department of Energy Grant No. DE-SC0009937 and by the World Premier International Research Center Initiative
(WPI Initiative), MEXT, Japan.  A.K. thanks Aspen Center for Physics for hospitality. 
S.N. is supported by JSPS (Japan Society for the Promotion of Science):
No.25610056, 26287056, and supported by MEXT (Ministry of Education, Culture, Sports, Science and Technology): No.26105521.
Y.Z.F. is supported by 973 Program of China under Grants No. 2013CB837000 and No. 11525313. B.B.Z. acknowledges the support by NASA TBD.

\end{document}